\documentclass[10pt]{article}
\usepackage{amsmath,amsthm,amscd,amssymb}

\newtheorem{thm}{Theorem}[section]
\newtheorem{prop}{Proposition}[section]
\newtheorem{lm}{Lemma}[section]
\newtheorem{ex}{Example}[section]

\newtheorem{cor}{Corollary}[section]

\title{On a linear code from a configuration of lines on the affine plane}
\author{Ken-ichi SUGIYAMA\thanks{Cooresponding address : Department of
Mathematics and Informatics, Faculty of Science, Chiba University, 1-33
Yayoi-cho Inage-ku, Chiba 263-8522, Japan. e-mail address : sugiyama@math.s.chiba-u.ac.jp}\\
Department of Mathematics and Informatics, Faculty of Science\\
Chiba University, Japan.}

\begin{document}
\maketitle
\begin{abstract}
We will show how to make a linear code from a configuration of affine
 lines in general position and a suitable set of rational points on
 it. The number of rational points on our singular curve is beyond the
 Weil bound and their coordinates are
 quite easy to compute. We will show a new decoding procedure
 which originates from the configuration. It is expected our method
 may correct errors less than almost the minimal distance {\it itself}, not
 the half of it.  
\par\vspace{5pt}
\end{abstract}

\section{Introduction}
In order to make a linear code from a nonsingular projective curve
 defined over a finite field ${\mathbb F}_q$ ($q$ is a power of a prime
 $p$), it should have many rational points. Let $X$ be such a
 curve of genus $g$ and $P=\{P_1,\cdots,P_N\}$ a set of rational points
 on it. We choose an effective divisor $D$ defined over ${\mathbb F}_q$ whose
 support is disjoint from $P$ and satisfying
\[
 2g-2 < {\rm deg}(D) < N.
\]
Then the evaluation map 
\[
 {\mathcal L}(D)\stackrel{\alpha}\to {\mathbb F}_q^N
\]
\[
\alpha(f)=(f(P_1),\cdots,f(P_N)), 
\]
embeds the linear system ${\mathcal L}(D)$ into ${\mathbb F}_q^N$ and
let $C(X,D)$ be its image. This is the basic construction of a linear
code using the algebraic geometry. It is known that its dimension is ${\rm
deg}(D)-g+1$ and the minimal distance is greater than or equal to
$N-{\rm deg}(D)$. (\cite{Lint}, \S 10.6)\\

But due to Weil, the number of rational points is bounded by
\[
 1+q+2g\sqrt{q},
\]
from above.
Even though one finds a good curve which attains the Weil bound, it is
quite hard to write down its equation explicitly and so is to obtain the
coordinates of rational points.\\

In \cite{Shibaki-Sugiyama}, in order to overcome this difficulty, we have proposed to use a configuration of affine lines on the affine
plane and have studied its general
properties. In this note we will show a explicit construction of a
good linear code from a certain configuration and a suitable set of
rational points on it. We will also investigate its property in detail
and will discuss a new
decoding procedure, which seems to be quite effective.\\

Let $\{L_1,\cdots,L_n\}$ be affine lines in a general
position on the affine $(x,y)$-plane which are defined over ${\mathbb
F}_q$, namely the intersection of every three of them is empty. Let
${\mathcal I}$ be the set of their intersection and we fix a positive
integer $m$. We take mutually distinct ${\mathbb F}_q$-rational points
$\{P_{i1},\cdots, P_{im}\}$ on $L_i$ disjoint from ${\mathcal I}$. Let
$d$ be a positve integer less than both of $m$ and $n$ and we put
\[
 {\mathcal F}_d=\{\sum_{i+j\leq d}a_{ij}x^iy^j\,|\, \,a_{ij}\in {\mathbb F}_q\}.
\]
Then the evaluation map embeds ${\mathcal F}_d$ into the space of $(n,m)$-matrices:
\[
 {\mathcal F}_d\stackrel{e}\to M_{nm}({\mathbb F}_q),\quad e(f)=(f(P_{ij}))_{ij},
\]
and its image is our linear code. The generating matrix can be
explicitly computed to be

\[
\left(
\begin{array}{ccccccc}
1& x(P_1) & y(P_1)& \cdots & x(P_1)^d & \cdots & y(P_1)^d\\
\vdots & \vdots & \vdots & \vdots & \vdots & \vdots & \vdots \\
1& x(P_{nm}) & y(P_{nm})& \cdots & x(P_{nm})^d & \cdots & y(P_{nm})^d
\end{array}
\right).
\]

The dimension of the code is
$\frac{(d+2)(d+1)}{2}$ and the minimal distance is greater than or equal
to $n(m-d)$ (resp. $m(n-d)$) if $m>n$ (resp. $n>m$). Moreover we have a new
decoding procedure which originates from the configuration. It is expected to correct errors less than
$n(m-d)$ or $m(n-d)$ if $m>n$ or $n>m$, respectively. \\

Here is an example to make $\{P_{ij}\}_{ij}$. Let
$\{L_1,\cdots,L_n,M_1,\cdots,M_m\}$ be affine lines in a general position and take $P_{ij}$ as the
intersection of $L_i$ and $M_j$. Then the minimal distance of
the code obtained from $\{L_1,\cdots,L_n\}$ and $\{P_{ij}\}_{ij}$ coincides with
$n(m-d)$ or $m(n-d)$ if $m>n$ or $n>m$, respectively. Therefore our decoding method may correct errors up to the
minimal distance {\it itself}, not the half of it.\\

By the Weil bound there are at most $1+q+(n-1)(n-2)\sqrt{q}$ rational
points on a nonsingular projective curve of degree $n$ in the projective
plane. But there are $nq-\frac{n(n-1)}{2}$ rational points on our
configration of affine lines. Thus, for fixed $n$, taking $q$ large enough, the number
of rational points on our curve is beyond the Weil bound. Moreover they are
quite easy to compute.

\section{Notation}
We will use the following notation throughout the paper.
\begin{itemize}
 \item For  a finite set $X$ its cardinality will be denoted
by $|X|$. 
\end{itemize}
Let $V$ be a vector space over  ${\mathbb F}_{q}$ of a finite
dimension. The function from $X$ to $V$ will be denoted by $V^{X}$,
which is a vector space of dimension $|X|{\rm dim}V$. 
\begin{ex} ${\mathbb F}_{q}^{\{1,\cdots,N\}}$ is isomorphic to ${\mathbb
 F}_q^{N}$ by the linear map:
\[
 \varphi(f)=(f(1),\cdots,f(N)),\quad f\in {\mathbb F}_{q}^{\{1,\cdots,N\}}.
\]
Using this we will identify them.
\end{ex}
Let $Y$ be a subset of $X$. Then there is a linear map
\[
 V^{X}\stackrel{r_{Y}}\to V^{Y},
\]
by restriction. The image of $v\in V^{X}$ will be denoted by $v_{Y}$.\\

Putting an arbitrary component whose index is not contained in $Y$ to be
zero, $V^{Y}$ may be considered as a subvector spave of $V^{X}$.
\begin{ex}
If one takes a subset $\Sigma$ of $\{1,\cdots,N\}$, ${\mathbb
 F}_q^{\Sigma}$ is identified with a subspace of ${\mathbb F}_q^{N}$
 defined as
\[
 \{(x_1,\cdots,x_N)\,|\, x_i=0\quad \mbox{if}\quad i\notin \Sigma\}.
\]
\end{ex}
 By
definition the restriction $r_Y$ to $V^Y$ is the identity.\\

Finally {\it the diagonal} $\Delta_{V^{X}}$ of $V^{X}$ is defined to be the set of
functions which take the same value at every element of $X$:
\[
 \Delta_{V^{X}}=\{f\in V^{X}\,|\, f(x)=f(x^{\prime})\quad \mbox{for
 any}\,x,\,x^{\prime}\in X\}.
\]
\section{A construction of a linear code}
Let us fix an $n$-tuple of affine
lines on the affine plane defined by a linear function $l_i$:
\[
 l_i=a_{i}x+b_{i}y+c_{i}, \quad a_{i},\,b_{i},\,c_{i}\in {\mathbb F}_{q}.
\] 
and $L_i$ the line defined by $l_i$. We assume that they are in a
general position. \\

If 
\[
 \left(
\begin{array}{ccc}
a_i & b_i & c_i\\
a_j & b_j & c_j\\
a_k & b_k & c_k\\
\end{array}
\right)
\]
is regular, the equation
\[
 \left\{
\begin{array}{c}
a_i x+ b_i y + c_i=0\\
a_j x+ b_j y + c_j=0\\
a_k x+ b_k y + c_k=0\\
\end{array}
\right.
\]
has no solution and we know the intersection of $\{L_i,\,L_j,\,L_k\}$
is empty. This observation shows the following lemma.
\begin{lm}
A family of affine lines
\[
 \{L_1\cdots,L_N\}
\] 
is in a general position if every $3\times 3$-minor of
\[
  \left(\begin{array}{ccc}
a_1 & b_1 & c_1\\
\vdots & \vdots & \vdots\\
a_N & b_N & c_N\\
\end{array}
\right)
\]
is regular.
\end{lm}

 The intersection of
$L_i$ and $L_j$, which is an ${\mathbb F}_{q}$ rational point, will be
denoted by $I_{ij}$. Let us choose mutually distinct $m$ points
$\{P_{i1},\cdots,P_{im}\}$ on $L_i$ which are 
 ${\mathbb F}_{q}$ rational and not contained in $\{I_{ij}\}_j$. The collection $\{P_{ij}\}_{i,j}$ will be denoted by ${\mathcal
 P}$. \\

Let ${\mathbb F}_q[x,y]$ be the polynomial ring of variables $x$ and $y$ with
 ${\mathbb F}_{q}$-coefficients. For a positive integer $d$ less than
 $m$ and $n$, we denote the subspace consisting of
 polynomials whose degrees are at most $d$ by ${\mathcal F}_d$. As a base of ${\mathcal F}_d$ we choose
\begin{equation}
 \{1,\,x,\,y,\cdots,x^d,\,\cdots,y^d\}.
\end{equation}
 In particular the dimension of ${\mathcal F}_d$ is
\[
 \delta=\frac{(d+2)(d+1)}{2}.
\]
Now we define {\it the evaluation map} 
\[
 {\mathcal F}_d\stackrel{e}\to {\mathbb F}_q^{\mathcal P}\simeq M_{nm}({\mathbb F}_{q})
\]
to be
\[
 e(f)=(f(P_{ij}))_{ij}.
\]
\begin{prop}
$e$ is injective.
\end{prop}
In order to prove the proposition we will prepare some notation.\\

By the lexicographic order, we arrange the indices of ${\mathcal P}$ as 
\begin{equation}
 ((1,1),\cdots,(1,m),\cdots,(n,1)\cdots,(n,m))=(1,\cdots,nm),
\end{equation}

which gives an identification between $M_{nm}({\mathbb F}_{q})$ and
${\mathbb F}_q^{nm}$. For a subset $\Sigma$ of ${\mathcal P}$, composing
with the restriction map, the evaluation map induces a linear map:
\[
 {\mathcal F}_d\stackrel{e_{\Sigma}}\to {\mathbb F}_q^{{\Sigma}}.
\]
 A subset
${\mathcal Q}$ of ${\mathcal P}$ will be mentioned {\it effective} if
there are distinct $(d+1)$-members $\{L_{q_1},\cdots,L_{q_{d+1}}\}$ of
$\{L_i\}_{1\leq i \leq n}$ such that the
cardinality of ${\mathcal Q}\cap L_{q_{\nu}}$ is $\nu$. In particular
$|{\mathcal Q}|$ is $\delta$. \\

The {\bf Proposition 3.1} immediately follows from the next proposition.
\begin{prop} For an effective set ${\mathcal Q}$,
\[
 {\mathcal F}_d\stackrel{e_{\mathcal Q}}\to {\mathbb F}_q^{{\mathcal Q}}.
\]
is an isomorphism.
\end{prop}
{\bf Proof.} Since the source and the target have the same dimension it is sufficient to
show $e_{\mathcal Q}$ is injective. Suppose $f\in {\mathcal F}_d$ satisfies
$e_{\mathcal Q}(f)=0$ and let $f_{\nu}$ be the restriction of $f$ to
$L_{q_\nu}$. Taking a linear parametrization of $L_{q_{\nu}}$, $f_{\nu}$
is a polynomial of one variable whose degree is at most
$d$. Let $\{Q_{\nu,1},\cdots,Q_{\nu,\nu}\}$ be the intersection of
${\mathcal Q}$ and $L_{q_\nu}$. We will show the following claim by an
induction for $\nu$.\\

{\bf Claim.}
The product $l_{q_{d+1}}\cdots l_{q_{d+1-\nu}}$ divides $f$ for $0\leq
\nu \leq d$.\\

For $\nu=0$ the assumption implies that $f_{d+1}$ vanishes at
mutually distinct $(d+1)$-points $\{Q_{d+1,1},\cdots,Q_{d+1,d+1}\}$.
Therefore $f_{d+1}$ vanishes because its degree is at most
$d$. This implies that $l_{q_{d+1}}$ divides $f$. \\

Let us assume the claim is
true for $\nu=i$. We know $f_{d-i}$ vanishes at $\{Q_{d-i,1},\cdots,Q_{d-i,d-i}\}$ by the assumption.
Moreover since $l_{q_{d+1}}\cdots l_{q_{d+1-i}}$ divides $f$, $f_{d-i}$ also vanishes on the
intersection of 
$L_{q_{d+1}}\cup \cdots \cup L_{q_{d+1-i}}$ and $L_{q_{d-i}}$, which is
$\{I_{q_{d-i,d-i+1}},\cdots, I_{q_{d-i,d+1}}\}$. Therefore $f_{d-i}$
vanishes at mutually distinct $(d+1)$-points and is zero by the reason of degree. This
implies that $f$ is divided by $l_{q_{d-i}}$. \\

The claim shows that $l_{q_{d+1}}\cdots l_{q_{1}}$ divides $f$, but
since the degree $f$ is at most $d$, it should be zero.
\begin{flushright}
$\Box$
\end{flushright}
We will consider the image of the evaluation map $e({\mathcal F}_d)$ as
a linear code.\\
 
Using the base (1) and the lexicographic order (2) the evaluation map has the following matrix representation:
\[
E=\left(
\begin{array}{ccccccc}
1& x(P_1) & y(P_1)& \cdots & x(P_1)^d & \cdots & y(P_1)^d\\
\vdots & \vdots & \vdots & \vdots & \vdots & \vdots & \vdots \\
1& x(P_{nm}) & y(P_{nm})& \cdots & x(P_{nm})^d & \cdots & y(P_{nm})^d
\end{array}
\right).
\]
It is nothing but the generating matrix of the code. 
\section{How to decode a message}
Let ${\mathcal E}_{n,m,d}$ be the family of effective sets. Its
cardinality is computed to be 
\[
 |{\mathcal E}_{n,m,d}|=\prod_{i=0}^{d}(n-i)\cdot 
\left(
\begin{array}{c}
m\\
d+1-i
\end{array}
\right).
\] 
For an effective set
${\mathcal Q}\in {\mathcal E}_{n,m,d}$, extracting the corresponding row vectors from $E$, we
obtain its $\delta\times \delta$-minor $E_{\mathcal Q}$. Then {\bf Proposition
3.2} shows it is a regular matrix. Using column vectors:
\[
 {\bf x}=
\left(\begin{array}{c}
1\\
x\\
y\\
\vdots\\
x^d\\
\vdots\\
y^d
\end{array}
\right)
,\quad
{\bf a}=
\left(\begin{array}{c}
a_{00}\\
a_{10}\\
a_{01}\\
\vdots\\
a_{d0}\\
\vdots\\
a_{0d}
\end{array}
\right),
\]
\[
 f=\sum_{i+j\leq d}a_{ij}x^iy^j \in {\mathcal F}_d
\]
has an expression
\[
 f={\bf x}^t\cdot {\bf a}.
\]
Moreover the image of evaluation map of $f$ is given by
\[
 e(f)=E\cdot {\bf a}.
\]
Now extracting components from both side whose indices are contained in
${\mathcal Q}$, we obtain
\[
 e(f)_{\mathcal Q}=E_{\mathcal Q}\cdot {\bf a},
\]
and
\[
 {\bf a}=E_{\mathcal Q}^{-1}\cdot e(f)_{\mathcal Q}.
\]
Thus we have proved
\begin{prop}
For an element ${\bf c}\in e({\mathcal F}_d)$, the vector $(E_{\mathcal
 Q}^{-1}\cdot {\bf c}_{\mathcal Q})_{\mathcal Q}$ is contained in the
 diagonal $\Delta_{({\mathbb F}_q^{\delta})^{{\mathcal E}_{n,m,d}}}$ of $({\mathbb F}_q^{\delta})^{{\mathcal E}_{n,m,d}}$. Moreover
 choosing arbitrary ${\mathcal Q}\in {\mathcal E}_{n,m,d}$, we have
\[
 {\bf c}=E\cdot E_{\mathcal Q}^{-1}\cdot {\bf c}_{\mathcal Q}. 
\]

\end{prop}
Let ${\bf c}=E\cdot {\bf a}$ be an element of the code.
We choose an arbitrary {\it error vector} ${\bf e}\in{\mathbb
F}_q^{nm}$ and set
\[
 {\bf c}^{\prime}={\bf c}+{\bf e},
\]
which may be considered as {\it a received message}.
We want to estimate the probability to hold
\begin{equation}
 E_{\mathcal Q}^{-1}{\bf c}^{\prime}_{\mathcal Q}=E_{{\mathcal Q}^{\prime}}^{-1}{\bf c}^{\prime}_{{\mathcal Q}^{\prime}}
\end{equation}
for ${\mathcal Q},\,{\mathcal Q}^{\prime}\in {\mathcal
E}_{n,m,d}$. If we put $\delta_{\mathcal Q}=E_{\mathcal Q}^{-1}{\bf
e}_{\mathcal Q}$ and $\delta_{{\mathcal Q}^{\prime}}=E_{{\mathcal
Q}^{\prime}}^{-1}{\bf e}_{{\mathcal Q}^{\prime}}$ respectively, {\bf
Prpposition 4.1} shows that (3) is equivalent to
\[
 \delta_{\mathcal Q}=\delta_{{\mathcal Q}^{\prime}}.
\]
The probability to hold this identity is
\[
 q^{-|{\mathcal Q}\ominus{\mathcal Q}^{\prime}|},
\]
where ${\mathcal Q}\ominus{\mathcal Q}^{\prime}={\mathcal
Q}\cup{\mathcal Q}^{\prime}\setminus ({\mathcal Q}\cap{\mathcal
Q}^{\prime})$.  This is very small if ${\mathcal Q}$ and ${\mathcal
Q}^{\prime}$ are different and if $q$ is sufficiently large. Therefore it is
expected that the following decoding procedure should be effective.\\

{\bf How to decode}\\

Take $q$ large and let ${\bf m}\in {\mathbb F}_q^{nm}$ be a received vector.
\begin{enumerate}
\item Compute ${\bf a}_{\mathcal Q}=E_{\mathcal Q}^{-1}{\bf m}_{\mathcal Q}$ for each effective set ${\mathcal Q}$.
\item If at least two of them coincide, search an element of $\{{\bf
      a}_{\mathcal Q}\}_{\mathcal Q}\in {\mathcal E}_{n,m,d}$ of the
      largest multiplicity. On the contrary if they are different each
      other, we think it is impossible to decode ${\bf m}$. 
\item Let ${\bf a}$ be the vector caluculated in {\bf Step 2}. Then the correct
      message should be $E\cdot{\bf a}$.  
\end{enumerate}
For distinct effective sets ${\mathcal
Q}$ and ${\mathcal Q}^{\prime}$, the previous estimate implies that if ${\bf m}_{\mathcal Q}$ or ${\bf m}_{{\mathcal
Q}^{\prime}}$ contains an error it should be quite rare that ${\bf
a}_{\mathcal Q}$ coincides with ${\bf a}_{{\mathcal
Q}^{\prime}}$. But if ${\bf m}$ contains too many errors, it may
happen that it is impossible to decode the message vector because every
two of $\{{\bf a}_{\mathcal Q}\}_{{\mathcal Q}\in \in {\mathcal
E}_{n,m,d}}$ may not coincide. In the next section we
will estimate the number of errors to be corrected.

\section{An estimate of the number of errors which may be corrected}
Let $R_{nm}$ be the following $n\times m$ rectangle with
grids:

\renewcommand{\arraystretch}{1.2}
\begin{center}
\begin{tabular}{|p{8mm}|p{8mm}|p{10mm}|p{8mm}|p{10mm}|}
 \hline
 $(1,1)$ & & $\cdots$ & & $(1,m)$ \\ \hline
$\vdots$ & & $\ddots$ & & $\vdots$ \\ \hline
$(n,1)$ & & $\cdots$ & & $(n,m)$ \\ \hline
\end{tabular}
\end{center}

Corresponding the grid $(i,j)$ to $P_{ij}$,  one may identify it with
${\mathcal P}=\{P_{ij}\}_{i,j}$. A subset $T$ of $R_{nm}$ will be mentioned as {\it a tableau} if it
satisfies the following condition:\\

If $(i,j)$ is contained in $T$, so is $(k,l)$ for $1\leq k \leq i$ and
$1\leq l \leq j$.\\

Here is a picture which illustrates the condition. $\heartsuit$ is a
grid contained in a tableau.

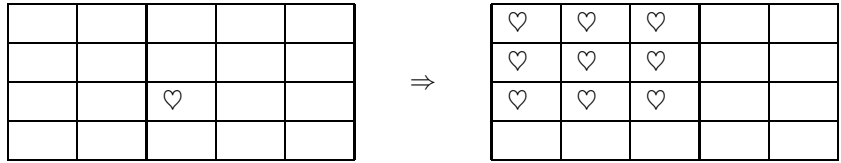
\begin{center}
\begin{tabular}{|p{5mm}|p{5mm}|p{5mm}|p{5mm}|p{5mm}|}
 \hline
  &  &  & &  \\ \hline
 & & & &  \\ \hline
 & &$\heartsuit$ & &  \\ \hline
 & &  & &  \\ \hline
\end{tabular}
\hspace{5mm}
$\Rightarrow$
\hspace{5mm}
\begin{tabular}{|p{5mm}|p{5mm}|p{5mm}|p{5mm}|p{5mm}|}
 \hline
 $\heartsuit$ &$\heartsuit$  &$\heartsuit$  & &  \\ \hline
 $\heartsuit$& $\heartsuit$&$\heartsuit$ & &  \\ \hline
 $\heartsuit$&$\heartsuit$ &$\heartsuit$ & &  \\ \hline
 & &  & &  \\ \hline
\end{tabular}
\end{center}

In general for a subset $\Sigma$ of $R_{nm}$, we denote the number of
grids contained in it by $\sigma(\Sigma)$.
\begin{ex}
\begin{enumerate}
\item (Regular tableau)
{\it The regular tableau} of size $l$ is 

\begin{center}
\begin{tabular}{|p{8mm}|p{8mm}|p{10mm}|p{8mm}|p{8mm}|}
 \hline
 $(1,1)$ & & $\cdots$ & & $(1,l)$ \\ \hline
 & & & \\ \cline{0-3}
  $\vdots$ &  & \\ \cline{0-2}
 & \\ \cline{0-1}
 $(l,1)$ \\ \cline{0-0}
\end{tabular},
\end{center}

which will be denoted by $R_l$. We have
\[
 \sigma(R_l)=\frac{l(l+1)}{2}.
\]
\item ($T_{k,l}$) The following tableau will be denoted by $T_{k,l}$:

\begin{center}
\begin{tabular}{|p{8mm}|p{8mm}|p{8mm}|p{8mm}|p{10mm}|}
 \hline
 $(1,1)$ &$\cdots$ & $(1,l)$ &$\cdots$ &$(1,m)$ \\ \hline
 $\vdots$ &$\ddots$ & $\vdots$ &$\ddots$ &$\vdots$ \\ \hline
 $(k,1)$ &$\cdots$ & $(k,l)$ &$\cdots$ &$(k,m)$ \\ \hline
  $\vdots$ & $\ddots$ & $\vdots$ \\ \cline{0-2}
 $(n,1)$ & $\cdots$ & $(n,l)$ \\ \cline{0-2}
\end{tabular}.
\end{center}

We have
\[
 \sigma(T_{k,l})=ln+km-kl.
\]
\end{enumerate}
\end{ex}
Let ${\mathcal C}$ be a subset of $R_{nm}$. We will consider a
sufficient condition so that it contains at least two effective
sets. Changing the numbering of lines and points, we may assume ${\mathcal
C}$ is a tableau.  
\begin{lm}
Let ${\mathcal C}$ be a tableau. If it contains $R_{d+1}$ and satisfies
\[
 \sigma({\mathcal C})>\sigma(R_{d+1})=\frac{(d+2)(d+1)}{2},
\]
it contains at least two effective sets.
\end{lm}
Since one can prove it by inspection, we only
show the simplest example of $d=2$:\\

If ${\mathcal C}$ is 
\begin{center}
\begin{tabular}{|p{5mm}|p{5mm}|p{5mm}|}
 \hline
 & & \\ \hline
 & \\ \cline{0-1}
  \\ \cline{0-0}
 \\ \cline{0-0}
\end{tabular},
\end{center}

it contains the following two effective sets which are marked by
$\heartsuit$:
\begin{center}
\begin{tabular}{|p{5mm}|p{5mm}|p{5mm}|}
 \hline
$\heartsuit$ &$\heartsuit$ &$\heartsuit$ \\ \hline
$\heartsuit$ &$\heartsuit$ \\ \cline{0-1}
 $\heartsuit$ \\ \cline{0-0}
 \\ \cline{0-0}
\end{tabular}
\end{center}

and

\begin{center}
\begin{tabular}{|p{5mm}|p{5mm}|p{5mm}|}
 \hline
 $\heartsuit$&$\heartsuit$ &$\heartsuit$ \\ \hline
 $\heartsuit$&$\heartsuit$ \\ \cline{0-1}
  \\ \cline{0-0}
$\heartsuit$ \\ \cline{0-0}
\end{tabular}.
\end{center}
\begin{lm} If a tableau $T$ does not contain $R_{d+1}$, we have
\[
 \sigma(T)\leq {\rm Max}\{f(k)\,|\,1\leq k \leq d+1\},
\]
where 
\[
 f(x)=x^2+(m-n-d-2)x+(n+1)(d+1)-m.
\] 
\end{lm}
{\bf Proof.} The assumption implies that there is $k$ with $1\leq k \leq
d+1$ such that the grid $(k,d+2-k)$ is not contained in $T$. Then by the
definition of a tableau, we see $T$ is contained in
$T_{k-1,d+1-k}$. Here is a picture which illustrates our situation:

\begin{center}
\begin{tabular}{|p{5mm}|p{5mm}|p{5mm}|p{5mm}|p{5mm}|p{5mm}|p{5mm}|}
 \hline
 $\heartsuit$ &$\heartsuit$ &$\heartsuit$ &$\heartsuit$ &$\heartsuit$ &$\heartsuit$ & \\ \hline
 $\heartsuit$&$\heartsuit$ & $\heartsuit$&$\heartsuit$ & & & \\ \cline{0-6}
 $\heartsuit$& $\heartsuit$& $\heartsuit$& $\spadesuit$ \\ \cline{0-3}
 $\heartsuit$&$\heartsuit$ & $\heartsuit$ \\ \cline{0-2}
 $\heartsuit$& $\heartsuit$ &  \\ \cline{0-2}
 $\heartsuit$&$\heartsuit$ &  \\ \cline{0-2} 
 & &  \\ \cline{0-2}
\end{tabular}
\end{center}

Here $\heartsuit$ are grids contained in $T$ and $\spadesuit$ is one at $(k,d+2-k)$. Hence we have
\begin{eqnarray*}
\sigma(T)&\leq& \sigma(T_{k-1,d+1-k})\\
&=& f(k)\\
&\leq & {\rm Max}\{f(k)\,|\,1\leq k \leq d+1\}.
\end{eqnarray*}

\begin{flushright}
$\Box$
\end{flushright}
\begin{cor}
Suppose that a tableau $T$ satisfies
\[
 \sigma(T)>{\rm Max}\{f(k)\,|\,1\leq k \leq d+1\},
\]
then it contains $R_{d+1}$.
\end{cor}
Notice that 
\[
 f(1)=nd,\quad f(d+1)=md.
\]
If $m$ or $n$ is greater than or equal to $d+2$ respectively, since $d$ is a
positive integer, we have
\begin{eqnarray*}
 f(d+1)&=& md\\
 &\geq &d(d+2)\\
& \geq & \frac{(d+2)(d+1)}{2}=\sigma(R_{d+1}),
\end{eqnarray*}
or
\[
 f(1)=nd \geq \sigma(R_{d+1}),
\]
respectively. This shows
\[
 {\rm Max}\{f(k)\,|\,1\leq k \leq d+1\} \geq \sigma(R_{d+1}).
\]
Combining {\bf Lemma 5.1} and {\bf Corollary 5.1} we obtain the
following theorem.
\begin{thm}
Suppose that $m$ or $n$ is greater than or equal to $d+2$. If a subset ${\mathcal C}$ of
 $R_{nm}$ satisfies
\[
 \sigma({\mathcal C})> {\rm Max}\{f(k)\,|\,1\leq k \leq d+1\},
\]
it contains at least two effective sets.
\end{thm}
From now we choose $m$ and $n$ so that one of them is greater than or
equal to $d+2$.\\

For a vector $\gamma=(\gamma_1,\cdots,\gamma_{nm})\in {\mathbb
F}_q^{nm}$ its {\it support} is defined to be
\[
 {\rm supp}(\gamma)=\{i\,|\,\gamma_{i}\neq 0\},
\]
and let $\nu(\gamma)$ be its cardinality. Let ${\bf m}$ be a received
vector.
It can be written as
\[
 {\bf m}={\bf c}+{\bf e},
\]
where ${\bf c}$ is an element of the code and ${\bf e}$ is an error. Let
${\mathcal C}$ be the complement of the support of ${\bf e}$.
{\bf Theorem 5.1} and {\bf Prposition 4.1} show, in the decoding procedure in the previous
section, if $\nu({\bf e})$ is less than $nm-{\rm Max}\{f(k)\,|\,1\leq k \leq d+1\}$,
at least two of $\{E_{\mathcal Q}^{-1}\cdot{\bf c}_{\mathcal Q}\}_{{\mathcal Q}\in{\mathcal
E}_{n,m,d}}$ coincide. Therefore it is expected that our decoding
procedure can correct errors less than $nm-{\rm Max}\{f(k)\,|\,1\leq k \leq d+1\}$.\\

Now we will estimate the minimal distance. Let ${\bf c}$ be an element
of the code and $T$ the complement of its support. As before we
may assume that $T$ is a tableau. {\bf Proposition 4.1} and {\bf
Corollary 5.1} show if $\sigma(T)=nm-\nu({\bf c})$ is greater than 
\[
 {\rm Max}\{f(k)\,|\,1\leq k \leq d+1\},
\]
${\bf c}$ should be zero. Thus we
know 
\[
 \nu({\bf c})\geq nm-{\rm Max}\{f(k)\,|\,1\leq k \leq d+1\}
\]
for every nonzero code vector ${\bf c}$, which implies the minimal
distance of $e({\mathcal F}_d)$ is greater than or equal to $nm-{\rm
Max}\{f(k)\,|\,1\leq k \leq d+1\}$. \\

Notice that, choosing ${\mathcal P}$ suitably, it is possible to construct a code
whose minimal distance is just $nm-{\rm Max}\{f(k)\,|\,1\leq k \leq
d+1\}$.\\

 In fact let $\{L_1,\cdots,L_n,M_{1},\cdots,M_{m}\}$ be a family
of affine lines in a general position which are defined by linear
functions whose coefficients are
in ${\mathbb F}_q$,
$\{l_1,\cdots,l_n,m_{1},\cdots,m_{m}\}$, respectively. Let $P_{ij}$ be the intersection of $L_i$ and $M_j$.
Suppose ${\rm Max}\{f(k)\,|\,1\leq k \leq
d+1\}$ is obtained at
$k=k_0$. If we take a polynomial $p$ of degree $d$ to be
\[
 p=\prod_{i=1}^{k_0-1}l_i \cdot \prod_{j=1}^{d+1-k_0}m_j,
\]
it is easy to see the complement of the support of $e(p)$ is
$T_{k_0-1,d+1-k_0}$. Thus we have
\begin{eqnarray*}
 \nu(e(p))&=&nm-\sigma(T_{k_0-1,d+1-k_0})\\
&=&nm-{\rm Max}\{f(k)\,|\,1\leq k \leq d+1\}.
\end{eqnarray*}
Here are some examples.
\begin{ex}
\begin{enumerate}
\item Suppose $m$ is greater than $n$. Then it is easy to see that
      $f(1)=nd$ is the maximum. Therefore it is expected that our
      decoding procedure may correct errors less than $n(m-d)$.
\item On the contrary suppose $n$ is greater than $m$. Then $f(d+1)=md$
      is the mamimum and it is expected that our decoding procedure may correct errors less than $m(n-d)$.
\end{enumerate}

\end{ex}

{\bf Acknowledgements}
The author is partially supported by the Grand-in-Aid for Scientific
Research (C) No.18540203, the Ministry of Eduvation, Culture, Sports,
Science and Technology, Japan.

\end{document}